\documentclass[journal]{IEEEtran}
\usepackage[utf8]{inputenc}
\def\BibTeX{{\rm B\kern-.05em{\sc i\kern-.025em b}\kern-.08em
    T\kern-.1667em\lower.7ex\hbox{E}\kern-.125emX}}

\usepackage{graphicx}
\usepackage{color}
\usepackage{cite}
\usepackage{amsmath}
\usepackage[caption=false,font=footnotesize]{subfig} 
\usepackage{amssymb}
\usepackage{comment}
\usepackage{mathtools}
\usepackage{tabulary}
\usepackage{acronym}
\usepackage{upgreek}
\usepackage{algorithm}
\usepackage{algpseudocode}
\usepackage{placeins}
\usepackage[super]{nth}
\usepackage{dsfont}
\usepackage{hyperref}
\usepackage{balance}

\newcommand{\tochange}[1]{#1}

\title{Metasurfaces-Integrated Wireless Neural Networks for Lightweight Over-The-Air Edge Inference}

\author{Kyriakos Stylianopoulos,~\IEEEmembership{Graduate~Student~Member,~IEEE}, Mario Edoardo Pandolfo,\\ Paolo Di Lorenzo,~\IEEEmembership{Senior~Member,~IEEE}, and George C. Alexandropoulos,~\IEEEmembership{Senior~Member,~IEEE}
\thanks{This work has been supported by the SNS JU projects 6G-DISAC and 6G-GOALS under the EU's Horizon Europe research and innovation program under Grant Agreements numbers 101139130 and 101139232, respectively.}
\thanks{K. Stylianopoulos and G. C. Alexandropoulos are with the Department of
Informatics and Telecommunications, National and Kapodistrian University of
Athens, 16122 Athens, Greece (e-mails: {kstylianop, alexandg}@di.uoa.gr).}
\thanks{M. E. Pandolfo and P. Di Lorenzo are with the National Inter-University Consortium for Telecommunications (CNIT), Parma, Italy.
M. E. Pandolfo is also with the DIAG Department, Sapienza University of Rome, via Ariosto 25, Rome, Italy.
P. Di Lorenzo is also with the DIET Department, Sapienza University of Rome, Via Eudossiana 18, Rome, Italy.
(e-mails: \{marioedoardo.pandolfo, paolo.dilorenzo\}@uniroma1.it).}
}

\begin{document}
\maketitle
\begin{abstract}
The upcoming sixth Generation (6G) of wireless networks envisions ultra-low latency and energy efficient Edge Inference (EI) for diverse Internet of Things (IoT) applications. However, traditional digital hardware for machine learning is power intensive, motivating the need for alternative computation paradigms. Over-The-Air (OTA) computation is regarded as an emerging transformative approach assigning the wireless channel to actively perform computational tasks. This article introduces the concept of Metasurfaces-Integrated Neural Networks (MINNs), a physical-layer-enabled deep learning framework that leverages programmable multi-layer metasurface structures and Multiple-Input Multiple-Output (MIMO) channels to realize computational layers in the wave propagation domain. The MINN system is conceptualized as three modules: Encoder, Channel (uncontrollable propagation features and metasurfaces), and Decoder. The first and last modules, realized respectively at the multi-antenna transmitter and receiver, consist of conventional digital or purposely designed analog Deep Neural Network (DNN) layers, and the metasurfaces responses of the Channel module are optimized alongside all modules as trainable weights. 
This architecture enables computation offloading into the end-to-end physical layer, flexibly among its constituent modules, achieving performance comparable to fully digital DNNs while significantly reducing power consumption. The training of the MINN framework, two representative variations, and performance results for indicative applications are presented, highlighting the potential of MINNs as a lightweight and sustainable solution for future EI-enabled wireless systems. The article is concluded with a list of open challenges and promising research directions. 
\end{abstract}

\section{Introduction}\label{sec:introduction}
Massive machine-type communications constitute one of the core use cases of fifth Generation (5G) networks, promoting the Internet of Things (IoT) paradigm. In the upcoming sixth Generation (6G), and beyond, ultra-low latency and energy efficient Device-to-Device (D2D) wireless links are envisioned, necessitating innovations toward power- and cost-efficient PHYsical (PHY) layer components, combined with unprecedented advancements in information processing algorithms and respective applications.

To deal with the anticipated amount of IoT-generated data, including Radio-Frequency (RF) signals intended for positioning and sensing, processing at the network edge is essential. To this end, 6G is expected to adopt a cross-layer design, transcending the hard distinctions between user plane and PHY layer. Additionally, the trend towards data-driven applications brings forth the role of Edge Inference (EI) and goal-oriented communications. \tochange{In the latter paradigm, the Transmitter (TX) encodes and sends data to the Receiver (RX), which does not aim to perfectly reconstruct them, but rather to extract information relevant to some computational task the network is designed to perform. EI constitutes a special case of goal-oriented communications focusing on facilitating the RX to understand a target feature of the transmitted signal that is not explicitly present in the original input data. To this end, a dataset of collected input-target pairs is leveraged to infer target values from past examples.} EI has benefits, first in terms of computations, since there is no need for RX to accurately reconstruct the input data, but also in terms of communication resources, since TX may choose encodings that preserve only information relevant to the target feature.

Machine Learning (ML) applications for IoT implemented at the network edge are gaining momentum in the landscape of 6G as an alternative paradigm of information processing. ML tools facilitate identification of patterns from data capturing precise characteristics of the intended deployment environment, which may not be accurately described by model-based approaches due to their unrealistic assumptions. Additionally, ML algorithms incur most of their computational complexity during training, which may take place offline at an earlier stage, offering low latency computations during deployment. On the contrary, the main cost associated with ML, especially for the EI domain, is that of hardware complexity. Parallel processing units are predominantly utilized to efficiently execute computations involved in Deep artificial Neural Networks (DNNs), resulting in substantial power consumption increases.

A transformative idea lately gaining traction posits that communications computational tasks need not be confined solely to the transceivers~\cite{AXN23_SIM}. Capitalizing on the notion
of \textit{goal-oriented smart wireless environments}, including infrastructure such as programmable MetaSurfaces (MSs) (a.k.a. Reconfigurable Intelligent Surfaces (RISs)~\cite{AlexandropoulosRIS}),
capable of sensing and intentionally reconfiguring the signal propagation environment \tochange{through intelligent beam shaping}, the wireless channel itself can become an active part of the computational chain. In this context, the channel evolves from a passive propagation medium into a goal-driven computational entity, effectively handling part of the processing load traditionally carried out by digital hardware. By shaping wave transformations Over-the-Air (OTA) using passive, active, or hybrid analog operations, the goal-oriented smart wireless environments paradigm envisions executing portions of the feature extraction, compression, filtering, or interference management computations with sufficiently low energy consumption. This redistributes, or even removes, computational load away from conventional energy-hungry TX/RX components and into the wave propagation level, enabling more sustainable and efficient ML-enabled wireless systems as well as ML applications such as EI.  


In this article, we investigate the integration of programmable multi-layer MSs into wireless propagation environments as well as their joint optimization with transceiver RF hardware components to realize OTA computations analogous to those performed by DNNs, thereby enabling an effective computational framework for wireless ML applications. In this way, ML computations that traditionally take place in digital processors can be realized with OTA operations directly in the domain of ElectroMagnetic (EM) waves propagation. The role of the MS-enabled smart wireless medium is therefore highlighted as a computational entity within which appropriately designed Multiple-Input Multiple-Output (MIMO) systems can be treated End-to-End (E2E) as single DNNs leveraging digital-, analog-, and wave-domain-based layers. Developing such PHY-layer-enabled wireless ML systems has the potential to greatly reduce the complexity and requirements for lightweight devices, like IoT, to perform EI.
\begin{figure*}
    \centering
    \includegraphics[width=\linewidth]{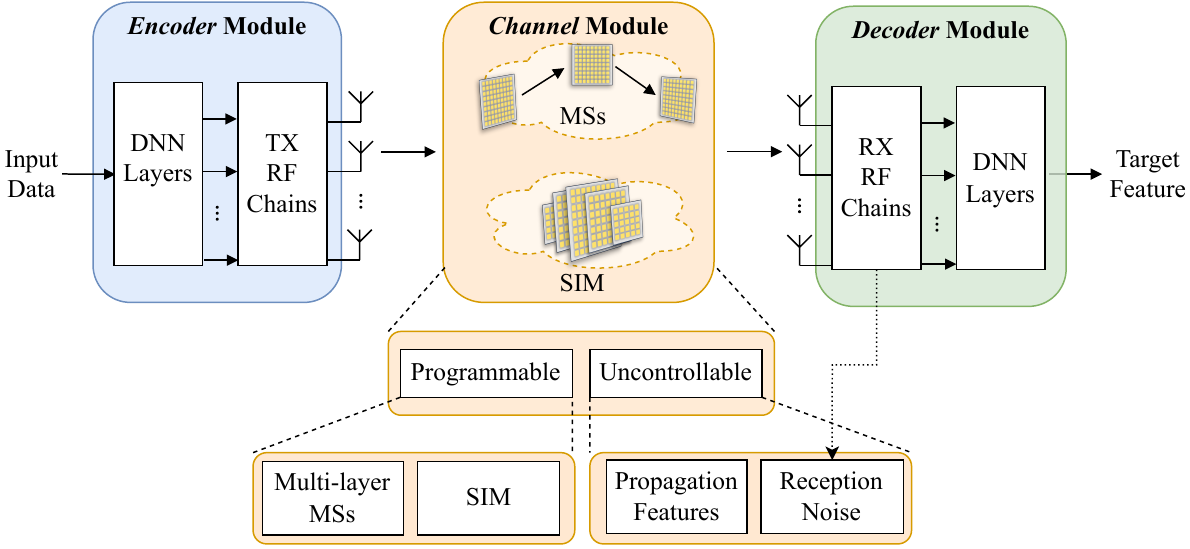}
    \caption{
    Conceptual architecture of the Metasurfaces-Integrated Neural Network (MINN) framework comprising three core modules. The {\em Encoder} and {\em Decoder} modules, which may incorporate neural network structures, are collocated respectively at the multi-antenna TX and RX nodes, while the {\em Channel} module performs OTA computations leveraging the fading coefficients of the wireless channel, the programmable EM responses of multi-layer MS structures constituting it, \tochange{as well as the properties of the RX thermal noise}. The E2E MIMO system is described as the composition of these three modules, therefore, the chain rule may be applied to compute the respective gradients and, consequently, optimize the heterogeneous DNN weights and constituent MS responses through gradient descent. It is noted that: \textit{i}) the physical MS device(s) enabling the reconfigurability of the signal propagation environment may be collocated at either in the TX or RX, instead of being placed in between them \tochange{(in this case, it is probable that the MS(s) affect only the portion of the signal components impinging on them)}, or in both; and \textit{i}) the {\em Encoder} and {\em Decoder} DNN components can be implemented either through conventional digital processors or equivalent analog computation units \tochange{(e.g., liquid state machines, memristors, and multi-layer MS structures, such as Stacked Intelligent Metasurfaces (SIM))}.}
    \label{fig:arch}
\end{figure*}

The remainder of this article is organized as follows.
Section~\ref{sec:d2nn} illustrates the basic principles behind MS-based ML and discusses their integration into wireless systems.
Section~\ref{sec:minn} presents the proposed Metasurfaces-Integrated Neural Network (MINN) framework, detailing training and deployment considerations, while Section~\ref{sec:variations} examines their multiple variations and applications. 
Section~\ref{sec:challenges} discusses open challenges with MINNs and outlines important research directions.
Finally, Section~\ref{sec:conclusion} concludes the article.

\section{Deep Diffractive Neural Networks}\label{sec:d2nn}
\tochange{This section discusses the fundamental principles of building MS-based DNNs and their integration to wireless systems.}

\subsection{Basic Principle}\label{sec:basic-principle}
The main technology enabling wave-domain implementation of DNNs is that of \tochange{Stacked Intelligent Metasurfaces (SIM)~\cite{AXN23_SIM}. Such multi-layer MS structures are composed of densely placed thin layers of diffractive MSs, each comprising multiple metamaterials with tunable EM responses, all managed by a controller. The overall SIM is typically supposed to be enclosed in absorbing material~\cite{XYN18_Diffractive_DNN}, and the propagation of the signal between elements of consecutive layers is governed by geometrical optics.} By purposely controlling the responses, one might perform particular operations on the signals that are forwarded from the first SIM layer. Since such operations are linear with respect to the impinging signal at every layer, and all elements of one layer contribute to the arriving signal at every element of the successor layer, this structure loosely resembles a \textit{fully-connected linear layer}, which is the fundamental component behind DNNs.
By exploiting this principle, Deep Diffractive Neural Networks (D\textsuperscript{2}NNs) may be materialized treating SIM responses as trainable weights~\cite{XYN18_Diffractive_DNN}. 

In the D\textsuperscript{2}NN context, the input data to the network must first be available in the RF domain.
This is readily done when the network is tasked with processing RF signals under sensing applications, which gives D\textsuperscript{2}NNs strong advantages to digital DNNs in terms of energy efficiency and latency, \tochange{since computationally lightweight devices are only needed}. In this case, no digital-to-analog conversions need to take place for data to be digitally processed. On the contrary, when D\textsuperscript{2}NNs are used for performing inference on digital input data, the transfer to the RF domain is of particular importance.
In~\cite{LML23_D2NN}, this conversion is achieved through programmable input layers at a SIM.
Each element of the input data vector is mapped to the response of a corresponding element of that layer using rudimentary techniques (e.g., amplitude modulation). Then, a beacon signal, typically from a single antenna, illuminates the back of the first layer to bootstrap the forward network pass.

To obtain D\textsuperscript{2}NN's output for inference applications, task-specific designs are necessary. In classification problems, where the output corresponds to the index of one or more predefined classes of which the input data are assumed to belong, signal receptors equal to the number of classes are placed after the final D\textsuperscript{2}NN layer. For single-class classification, the predicted class index corresponds to the index of the receptor with the higher observed signal strength. In regression problems, obtaining D\textsuperscript{2}NN's output is relatively less straightforward. It is possible to interpret the signal at receptors as being amplitude- or phase-modulated, however, extremely fine grained beams need to be implemented by the SIM to achieve desired accuracy; this is impractical with the available D\textsuperscript{2}NN hardware designs~\cite{LML23_D2NN}. Considering the training phase in particular, once the forward pass is performed and the output is converted through analog-to-digital converters, the result is compared to the expected target value for each of the training data instances, and the loss function is then digitally computed.
The backpropagation algorithm is then applied to determine the changes in the responses in each of the SIM elements and the process is repeated until convergence.

Once the data is converted to RF signals, D\textsuperscript{2}NNs perform computations at the speed of light, giving them a competitive advantage over digital DNNs.
Arguably, more important are the benefits in terms of power consumption. MSs usually comprise near-passive circuitry, such as varactors, requiring even nanoWatts to operate~\cite{AlexandropoulosRIS}.
This is a tremendous energy efficiency improvement compared to standard DNN processors (particularly, Graphics Processing Units (GPUs)), which consume hundreds of Watts for inference.
Even more so, once the desired responses of constituent metamaterials are determined (presumably in simulating software), the MSs may be fabricated to be completely passive. In such cases, the only power consuming components of D\textsuperscript{2}NNs  are therefore the feeding antenna and output receptors, all of which may operate under very low power settings in controlled wireless environments where signal attenuation and multipath effects are negligible. It is noted that, for applications where the input data are already in the RF domain, such as those that fall under communications, localization, and/or sensing, the elimination of the need of digital-to-analog conversion, and its associated power and latency overheads, constitutes D\textsuperscript{2}NNs an ideal candidate for ``exotic'' analog DNN hardware, \tochange{enabling lightweight ML applications, such as EI}.

\subsection{Integration in Wireless Systems}\label{sec:integration-d2nn}
Despite their analog nature, D\textsuperscript{2}NNs have been predominantly developed as DNN hardware accelerators for general inference problems~\cite{LML23_D2NN}. To materialize the promised advantages under EI applications, the integration of SIM-based computing within wireless communication infrastructure requires further investigation.
In fact, the current design of D\textsuperscript{2}NNs is largely incompatible with existing and future wireless system stacks. 

One of the main considerations is the digital-to-analog conversion of input data. Encoding each element of this data (e.g., image pixel) directly as a pre-mapped response of a programmable first SIM layer might be severely constraining for practical applications, since: \textit{i}) the power consumption attributed to the SIM controller might be significant; \textit{ii}) it is difficult to obtain sufficient precision in programmable MS responses to accurately encode input data; and \textit{iii}) the size of the first SIM layer grows with the dimensionality of the data. Besides, current D\textsuperscript{2}NN mode of operation does not utilize the capabilities of contemporary wireless systems.
In particular, MIMO systems leverage transmission over multiple antennas to achieve spatial multiplexing or beamforming, thus, providing additional degrees of freedom for feeding input data to the SIM device.
Under this prism, standard PHY operations, such as source encoding, modulation, and precoding/combining, could be exploited to integrate D\textsuperscript{2}NNs in MIMO systems.

\tochange{Another important consideration of SIM-based computing for EI is the wireless channel. D\textsuperscript{2}NNs have been originally developed for almost free-space signal propagation conditions with high Signal-to-Noise Ratio (SNR) levels~\cite{LML23_D2NN}. However, in practical wireless systems, small- and large-scale fading effects resulting in fluctuating SNR levels are present.} For these systems, a SIM can play a dual role: their learned responses can perform successful inference, akin to DNN layers, while also adapting to channel conditions. It is important to highlight that the wireless channel needs not be treated as a source of undesirable behavior; it can be employed as an additional means of computation, following the OTA computing paradigm, \tochange{where the superimposition of wireless signals is exploited to carry out computations~\cite{AirComp_Review}}.

\section{Metasurfaces-Integrated Neural Networks}\label{sec:minn}
\tochange{In this section, we elaborate on the MINN framework~\cite{Stylianopoulos_GO}, a generic MIMO wireless system setup operating E2E as a single DNN, integrating digital and/or analog neural network layers, beamforming operations, and D\textsuperscript{2}NN layers. This PHY-layer-enabled DNN framework consists of three core modules, as illustrated in Fig.~\ref{fig:arch}. Each module represents a physical entity implemented in a distinct system device (i.e., TX, RX, and multi-layer MS structures and/or SIM realizing the smart wireless environment), as detailed in the sequel.}

\tochange{\subsection{Overall Architecture}\label{sec:minn-arch}}
\tochange{As depicted in Fig.~\ref{fig:arch},} the TX operates the {\em Encoder} module, \tochange{which feeds the data this device possesses into their neural network} layers to directly output the signal that is to be transmitted. In that regard, TX is responsible for initial feature extraction, compression, and modulation, while accounting for error correction and other channel-related operations. When performing EI, the principles of joint source-channel coding are arguably more convenient to follow, suggesting that all of the above operations are performed implicitly through the hidden layers of the module, rather than devising separate functional blocks for each one \tochange{(this strategy is common under separated source-channel coding schemes of conventional communications)}. Regardless of the architecture of the module, the transmitted signal should be in a form that makes use of all degrees of freedom offered by modern MIMO systems (in the spatial, temporal, and frequency domains), and adheres to average or maximum power constraints. Three-dimensional complex-valued tensors can be used to represent the TX signal, with each element representing the baseband signal at a certain antenna, frequency bin, and time slot. In practice, the {\em Encoder} can be implemented through a conventional DNN, however, in EI applications involving lightweight devices with limited requirements, this module's complexity is a restrictive factor.

\tochange{The {\em Channel} module within} the MINN framework contains two components, \tochange{an uncontrollable and a programmable one}. \tochange{The former component consists of the typical wireless} channel itself \tochange{together with the thermal noise} at the RX side. \tochange{Wave propagation is well known to be subject to fading, thus, imposing linear effects on the transmitted signal, whereas thermal noise is commonly additive white Gaussian.} The second component concerns the \tochange{programmable MSs influencing the wireless TX-RX link}. \tochange{These devices constitute key features of the overall programmable MIMO channel, offering dynamically reconfigurable channel matrix coefficients.} Under the MINN framework, the responses of multi-layer MS structures and/or SIM are treated similar to trainable weights of typical DNN layers: they are optimized through a training process \tochange{with the goal to configure the overall} channel to perform \tochange{OTA} computations aiding inference. \tochange{In this way, ML computations, that would otherwise be performed via digital DNN layers, can be realized by the programmable MIMO channel, thereby, offloading either the TX, RX, or both.} This represents a paradigm shift over traditional communication systems, where the wireless channel is treated as a component of adverse effects motivating countermeasures at the transceivers. 
\tochange{Note that the precise form of computations that are attainable by the {\em Channel} module varies depending on the wireless environment as well as the capabilities of MSs/SIM. Most of state-of-the-art designs for the latter~\cite{XYN18_Diffractive_DNN,LML23_D2NN} offer computations maintaining the linear nature imposed by the channel response matrix, and only recently nonlinear implementations have started being considered~\cite{Stylianopoulos_MIMO_ELM,NL_ELM}, as will be discussed in the sequel.} 
\begin{figure*}[t]
    \subfloat[\tochange{MS layers each with} $8\times8$ metamaterials and SNR of $5$~dB.\label{fig:minn-clf-a}]{%
        \includegraphics[width=\columnwidth]{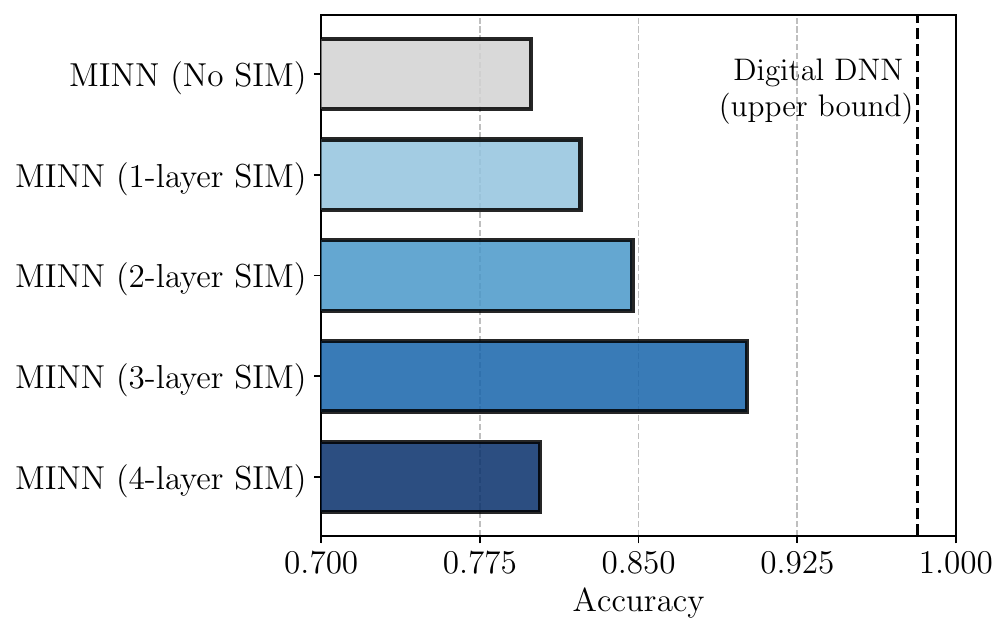}
    }
    \subfloat[\tochange{MS layers each with} $12\times12$ metamaterials and SNR of $10$~dB.\label{fig:minn-clf-b}]{%
        \includegraphics[width=\columnwidth]{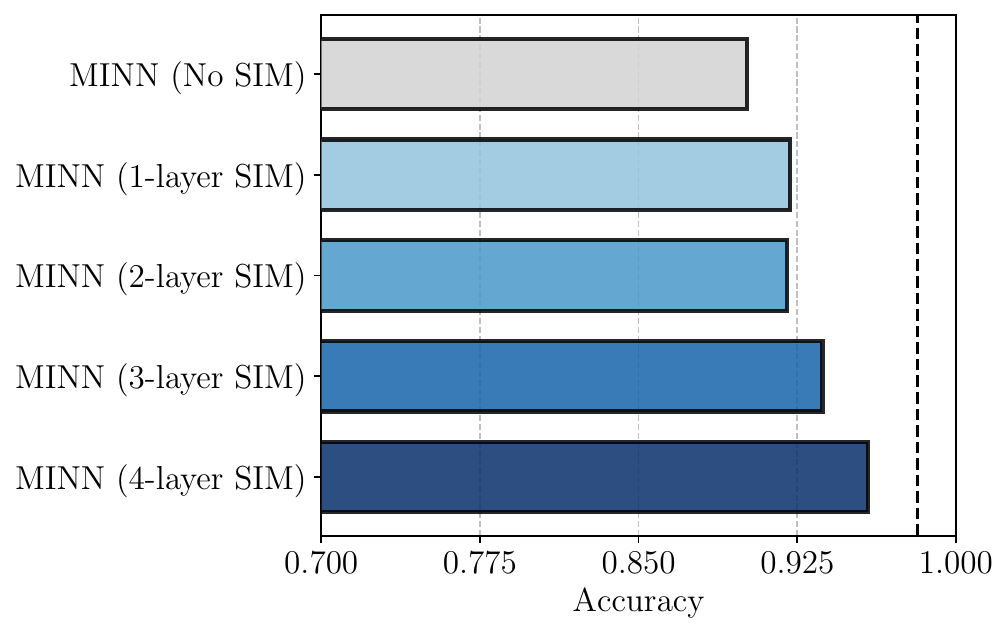}
    }
    \caption{Mean accuracy of different MINN versions for MNIST classification, \tochange{considering fixed SNR during training and inference. All simulated $4\times4$ MIMO system setups included a SIM positioned close to the TX at a distance corresponding to approximately 7.5\% of the total TX-RX distance, with its broadside perpendicular to the TX-RX line of sight, differing on the number of MS layers and the number of metamaterials per layer. A geometric channel with $10$ scatterers, yielding static fading conditions, has been considered.} Two convolutional layers followed by a linear layer were used at the TX digital DNN, while the RX digital DNN comprised three feedforward layers. The ``No \tochange{SIM}'' baseline refers to performing inference \tochange{with only the digital DNNs at the transceivers, i.e., without manipulating the wireless channel through any MS (a simplified MINN variation including only the uncontrollable component of the {\em Channel} module).} The ``Digital DNN'' benchmark refers to performing MNIST classification entirely on the TX with the same number of layers, without accounting for channel transmission, and is therefore an upper bound. 
    \tochange{All networks were trained for $50$ epochs irrespective of their size.} As observed, by increasing the number of \tochange{SIM} elements and the received SNR level, larger MINNs approach the Digital DNN bound, whereas the removal of the SIM is detrimental to the training process. \tochange{It is additionally demonstrated that, under lower SNRs, deeper MINN versions are not always more efficient since they suffer from higher signal attenuation through the SIM layers.}}
    \label{fig:minn-classification}
\end{figure*}

\tochange{As shown in Fig.~\ref{fig:arch}, the final module of the MINN framework is the {\em Decoder}, which is realized at the RX. The signal passing through the {\em Channel} module is collected therein, with the goal} to extract the embedded information and output the inference result. To achieve this, operations similar to channel equalization, demodulation, and source decoding are implicitly implemented. Crucially, the {\em Decoder} does not aim to reconstruct the exact form of the input signal, but rather to perform feature extraction on it. \tochange{Similar to the {\em Encoder}, realizing the {\em Decoder} through a conventional DNN entails increased complexity which may be prohibitive for certain inference applications through lightweight wireless devices.}

\tochange{\subsection{Training for Static and Dynamic MS Responses}\label{sec:training}}
The sequence of operations described above concern the forward pass of the \tochange{considered PHY-layer-enabled DNN framework}. The overall model can be expressed mathematically as a functional composition of the {\em Decoder}, {\em Channel}, and {\em Encoder} modules in that order~\cite{Stylianopoulos_GO}. This conceptual structure is therefore compatible with the backpropagation algorithm for DNN training, where the last layer/module has its trainable weights updated first through Stochastic Gradient Descent (SGD), and each preceding layer has its weights updated with respect to the gradient of its successor layer or module.

For the case of MINN \tochange{training on varying fading with stationary distributions}, this procedure can be detailed as follows. During training, each data instance is paired with a \tochange{different} random channel instance, and the forward pass under these channel conditions is performed to derive the network's output value. The error between that prediction and the known target value corresponding to the input data is measured with the help of an objective function (e.g., mean squared error or cross entropy). The gradient with respect to the last layer of the {\em Decoder} is then computed, and the error signals are backpropagated through this module, which are in turn forwarded to the {\em Channel} and {\em Encoder} modules. \tochange{Note that this training procedure implies that the data is stochastically independent from the channel conditions, which might not always guaranteed (e.g., when data arises from wireless sensors in target sensing scenarios, the presence of a target affects the channel).} Assuming training takes place in an accurate simulator, all DNN weight updates can be digitally computed and only the trained modules need to be deployed. \tochange{This implies that the MSs included at any of the MINN modules can be fabricated to implement the optimized responses in a fixed passive manner.} \tochange{Note that this procedure also treats the case of dynamic channel fading variation, and, as a result, the trained MINN can adapt to instantaneous channel changes as long as their statistics remain the same. This happens, of course,} at the expense of prolonged training times, since the effective dataset is the Cartesian product of all data and channel instances.

Another theoretical consideration is the effect \tochange{of random thermal noise in the {\em Channel} module.} Since this noise is additive, it does not impede the calculation of gradients. Moreover, noise whiteness guarantees that each observation of the received signal is an unbiased estimator of its noise-free version, a fact that ensures SGD convergence. Evidently, cases with low received SNRs require more extensive training, and, empirically, it is beneficial to pre-train MINNs first with data at high SNR values and, then, adapt to low SNR conditions through fine tuning and transfer learning techniques~\cite{Stylianopoulos_GO}. \tochange{A MNIST classification performance evaluation with the presented MINN architecture is demonstrated in Fig.~\ref{fig:minn-classification} for static fading conditions. As observed, in the high SNR regime and with sufficient number of MS elements, the performance of the presented PHY-layer-enabled DNN framework with optimized static SIM approaches that of fully digital DNNs.}

\tochange{The MINN framework also includes the option where the MS-based DNN layers, at any of the MINN modules, adapt their constituent responses at every channel coherent block. This option, however, necessitates a dedicated DNN at each module hosting MS(s), serving as the structure's dynamic response controller~\cite{Stylianopoulos_GO}. These controllers need to be trained in an E2E coordinated manner to map instantaneous channel observations to MS responses, a fact that increases the number of trainable DNN parameters and inter-module training control overhead.} Nevertheless, in highly dynamic wireless conditions with correlated fading, this MINN version with dynamic MS responses promises significant performance improvements. 

\tochange{\subsection{Two MINN Architecture Variations}\label{sec:arch-variations}}
\tochange{The MINN framework offers a large variety of potential implementations, leveraging the computing potential of either multi-layer MS structures, or MIMO systems with eXtremely Large (XL) antenna arrays, or both. The first all-MSs variation enables replacing digital DNN layers with wave-domain-based DNNs~\cite{GJZ24_SIM_TOC}, resulting in computationally lightweight TX and/or RX devices, whereas the second exploits the XL number of transformations offered by the wireless channel.}

\begin{figure*}[t]
    \centering
    \subfloat[\centering \tochange{An all-MSs MIMO system where all except the first and last MS-based DNN layers, placed respectively at the TX and RX, can be flexibly installed within the signal propagation environment or/and near the end devices.} \label{fig:sim_positions}]{%
       \makebox[\linewidth][c]{\includegraphics[width=0.8\linewidth]{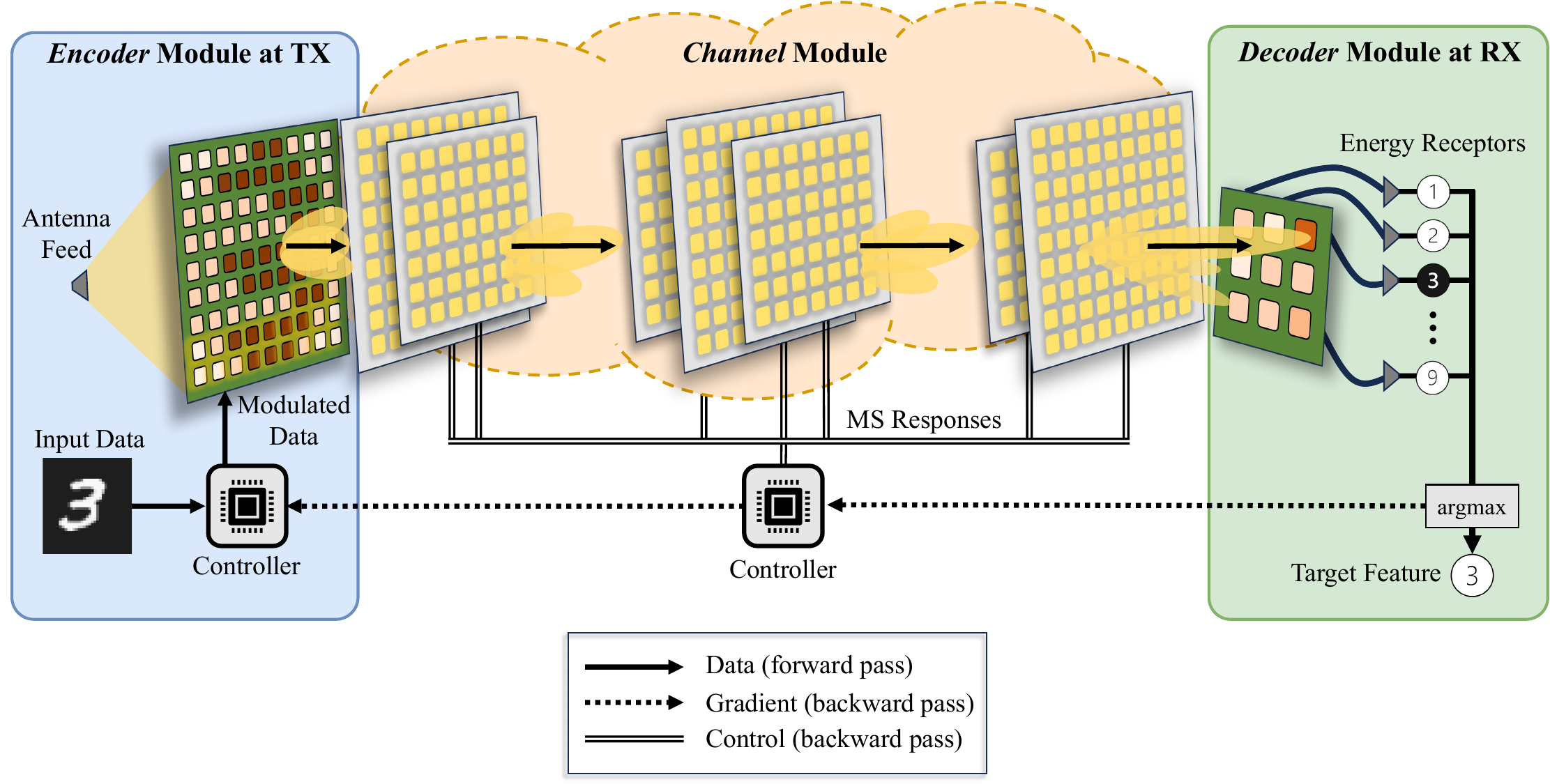}}
    }\\ 
    
    \subfloat[\tochange{An XL MIMO system with analog combining of the nonlinearly processed received signals, leveraged as a single hidden layer neural network.} \label{fig:elm_mimo}]{%
       \makebox[\linewidth][c]{\includegraphics[width=0.8\linewidth]{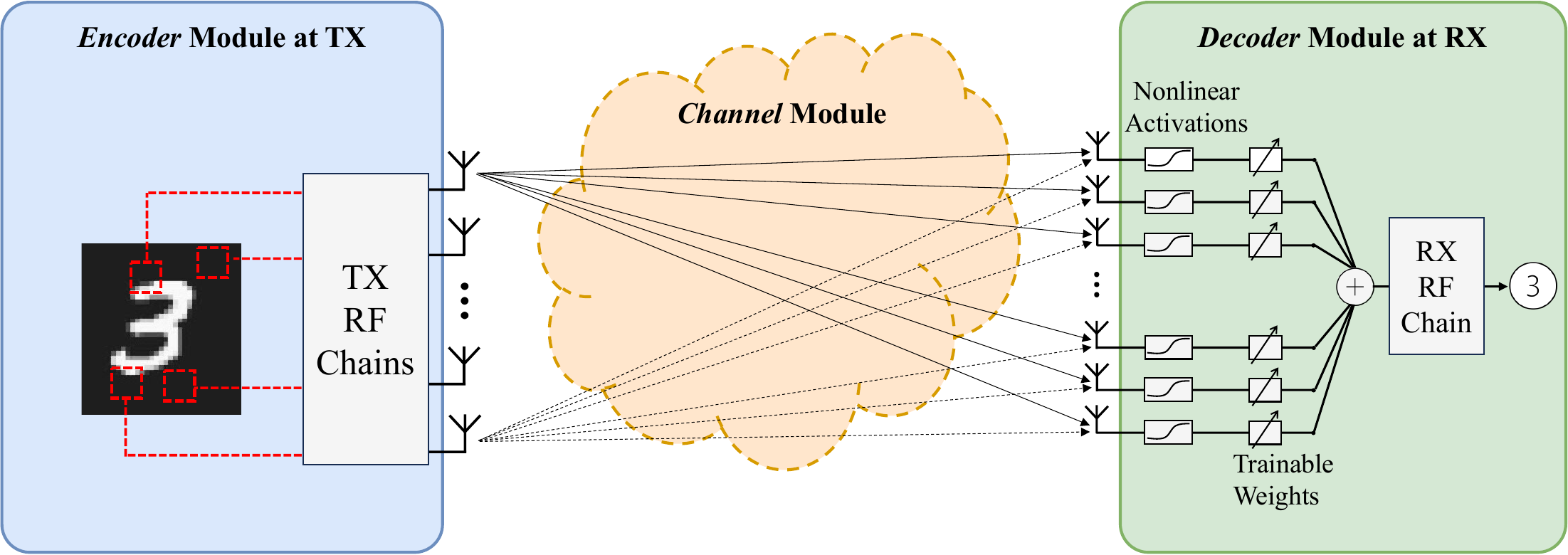}}
    }
    
    %
    \caption{\tochange{Two variations of the MINN architecture for the example of MNIST handwritten digit classification: (a) An all-MSs MIMO system where all three modules are primarily implemented by MSs. At the TX, the forward network pass initiates through a single antenna illuminating the first layer of the E2E wave-domain-based DNN, which has as many elements as the number of features within the input data (i.e., the number of MNIST image pixels). This data is encoded via the EM responses of this layer's diffractive metamaterials~\cite{LML23_D2NN}. At the RX, the final MS layer, constituting of ten fully absorbing metamaterials each followed by an energy detector (with each representing one of the possible MNIST digits), provides the output inferred digit. Although conceptually regarded as parts of the {\em Channel} module handling all OTA computations, the hidden diffractive MS layers may be flexibly distributed at all system physical devices. For example, some can be collocated with the transceiver devices and the remaining installed all together or in groups within the signal propagation environment. Alternatively, those diffractive MSs may be placed at one of the end devices and inside the MIMO channel, or solely within that channel, enabling lightweight transmissions, receptions or both. (b) An XL MIMO system where the {\em Encoder} and {\em Decoder} modules are respectively carried out by a multi-antenna TX and RX, and the {\em Channel} module is devoid of programmable devices.~\cite{Stylianopoulos_MIMO_ELM}.
    At the TX, each pixel of the input data is encoded and fed to a distinct antenna, while, at the RX, the signal received at each antenna is first fed to a nonlinear RF component, then, it is multiplied by a controllable weight, and, finally, all weighted analog outputs are combined to provide the output inferred digit. 
    This XL MIMO system capitalizes on the random transformations imposed by the uncontrollable {\em Channel} module on the feature signals before being superimposed at the RX antennas, operating as an OTA ELM.}}
    \label{fig:MS-positions}
\end{figure*}

\tochange{An example MINN architecture for MNIST handwritten digit classification based primarily on MSs at all three modules is depicted in Fig.~\ref{fig:MS-positions}(a). The first ({\em Encoder} module) and last ({\em Decoder} module) MS layers of the wave-domain-based E2E DNN are respectively implemented at the TX and RX, while the hidden MS layers constituting the {\em Channel} module are handled by the programmable MIMO channel~\cite{Stylianopoulos_GO}. The latter implies that the smart wireless environment may act as shared OTA computing resources that can be dynamically allocated to lightweight end devices for ML applications~\cite{GJZ24_SIM_TOC}. Under this viewpoint, future IoT networks may be designed to offer ``OTA computation as a service,'' in order to partially offload devices from computationally demanding D2D inference tasks.}

\begin{figure*}[t]
    \subfloat[Classification on Parkinson's and MNIST datasets.\label{fig:elm-a}]{%
        \includegraphics[width=\columnwidth]{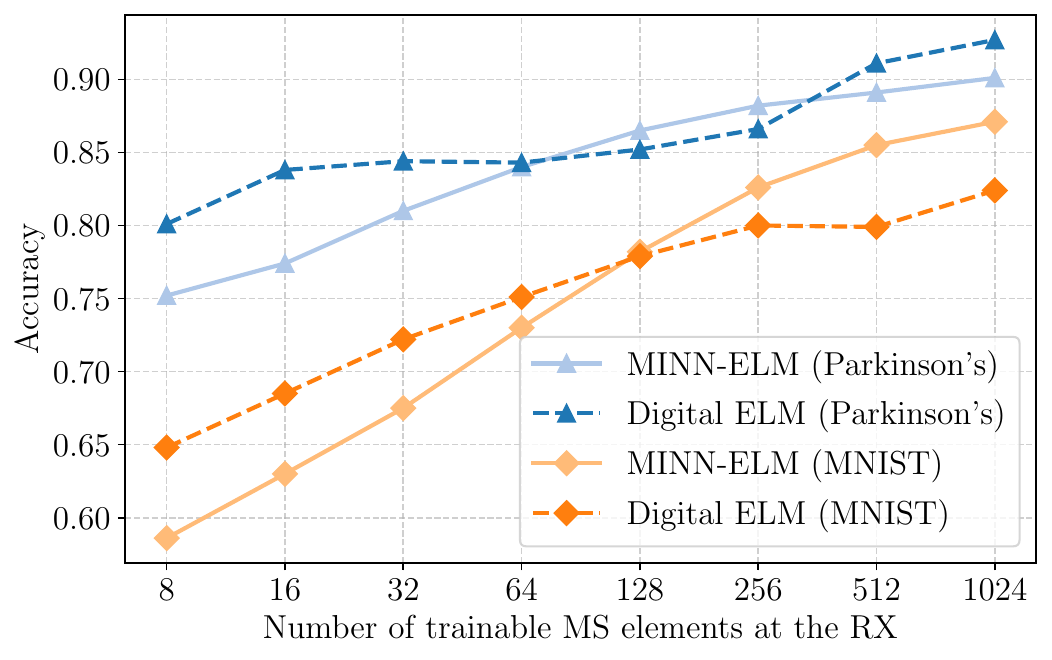}
    }
    \subfloat[Classification on the WBCD and SECOM datasets. \label{fig:elm-b}]{%
        \includegraphics[width=\columnwidth]{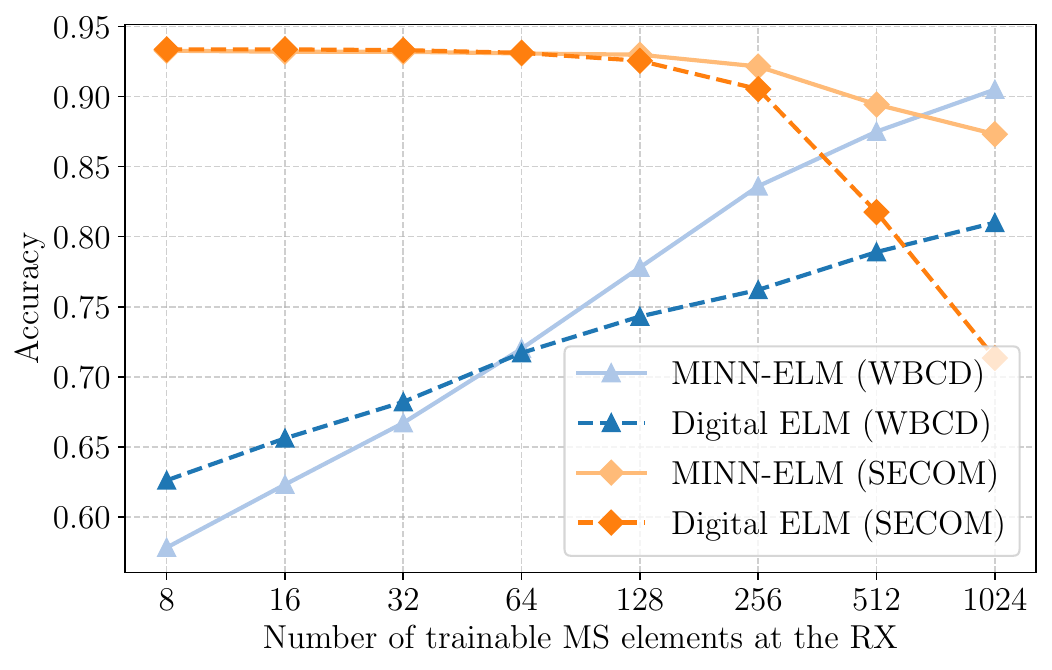}
    }
        \caption{Mean accuracy of the MINN-ELM variation in Fig.~\ref{fig:elm_mimo} over different binary classification datasets at the received SNR level of $25$~dB. \tochange{XL MIMO system setups with different numbers of TX antenna elements and MS sizes at the RX were simulated under a Rayleigh fading channel, which was treated as the random single hidden layer of the E2E wave-domain-based neural network.} Each constituent metamaterial of the last MS-based layer \tochange{at the RX} was designed to realize a cascade of a fixed nonlinear response (thus, acting as an activation function) followed by a tunable linear response (thus, acting as a trainable weight). The MINN-ELM training took place in closed form within each channel coherence block. \tochange{The number of TX antennas corresponded to the number of input features of the dataset used: $22$ for Parkinson's; $60$ (sub-sampled from $784$) for binary (even/odd) MNIST; $30$ for Wisconsin Breast Cancer Diagnosis (WBCD); and $20$ (sub-sampled from $590$) for Semiconductor Manufacturing (SECOM). The MINN-ELM classification performance has been compared with that of a fully digital ELM implementation, considering $200$ random initializations over four datasets.} \tochange{As observed, The MINN-ELM performs equally well to its digital counterpart in all scenarios. The performance increases as its approximation power is enhanced by increasing the number of MS elements at the RX, with the exception of the SECOM dataset which suffers from overfitting. In sufficiently XL MIMO conditions, MINN-ELM achieves} near-optimal classification for all datasets, approaching their asymptotic theoretical guarantees of universal approximation.}
    \label{fig:elm}
\end{figure*}

\tochange{Figure~\ref{fig:elm_mimo} illustrates an XL MIMO system operating as an Extreme Learning Machine (ELM)~\cite{Stylianopoulos_MIMO_ELM}. This single-hidden-layer E2E neural network paradigm has been recently shown to offer OTA digit classification leveraging the uncontrollable XL MIMO channel. In particular, the TX hosting the {\em Encoder} module performs analog modulation, encoding digit images at their XL antenna array. This first neural network layer is followed by the hidden layer formulated by the channel gain coefficients of the uncontrollable {\em Channel} module. The last layer constituting the {\em Decoder} module, realized at the XL RX, collects the received signals at each antenna element, feeding each of them to a nonlinear RF component. Then, it multiplies each of them with a controllable weight and, finally, all weighted analog outputs are combined to provide the output inferred digit. For the latter purpose, the RX may deploy an adequately designed conventional antenna array followed by RF circuitry~\cite{Stylianopoulos_MIMO_ELM} or a diode-based diffracting MS(s)~\cite{NL_ELM} to perform nonlinear activation. It was recently shown in~\cite{Stylianopoulos_MIMO_ELM} that, for static fading conditions, the combining weights of this MINN-ELM variation can be optimized in closed form, and may be quickly fine-tuned as fading changes gradually over time without relying on costly backpropagation. As demonstrated in Fig.~\ref{fig:elm}, MINN-ELMs with an XL MS at the RX can perform equally well to fully digital ELMs that ignore fading. In fact, it has been proven in~\cite{Stylianopoulos_MIMO_ELM} that MINN-ELMs are universal approximators (i.e., they may approximate any computable function based on available data) under the following conditions: \textit{i}) XL numbers of RX antennas; \textit{ii}) rich scattering conditions  (Rayleigh-like); and, crucially, \textit{iii}) nonlinear response at the MS elements. Overall, MINN-ELMs exploit random fading as linear projections of input data to an arbitrary, yet representationally rich space, and use the MS responses at RX as weights that promote nonlinear combinations of useful representations to the inference problem at hand.}

\tochange{\subsection{Example MINN Applications}\label{sec:variations}}

\begin{figure}[t]
    \centering
    \includegraphics[width=\linewidth]{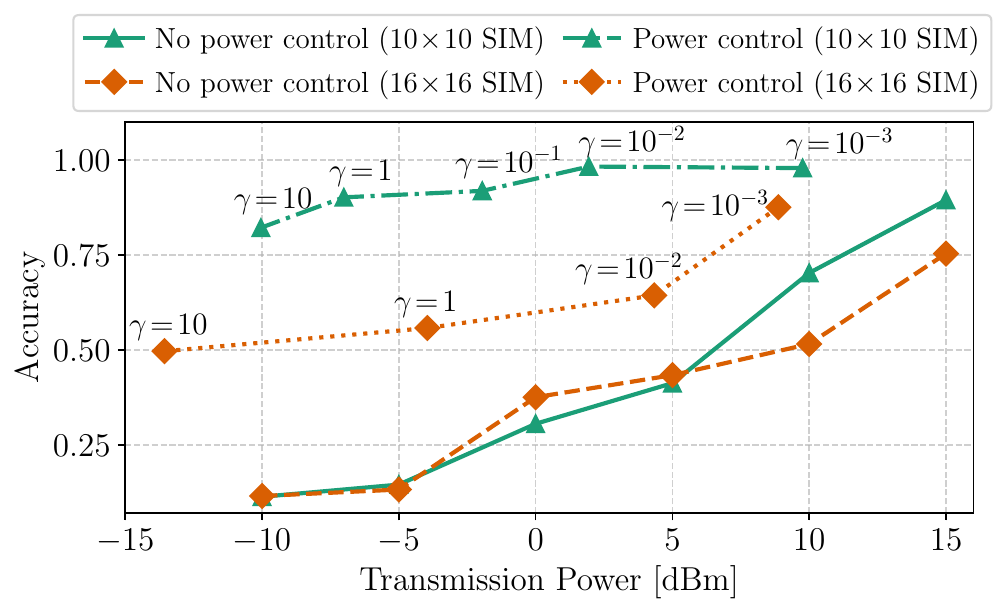}
    \caption{Mean accuracy of MINN with power control for MNIST classification versus the TX power level, compared with MINN trained under constant power budgets. \tochange{An $16 \times 8$ MIMO system setup with two different sizes of $4$-layer SIM, positioned close to the TX similar to Fig.~\ref{fig:minn-classification}, operating under a geometric channel with $15$ scatterers, was simulated. At each channel realization, the RX position was randomly sampled, resulting overall in dynamic fading conditions.} The penalty term $\gamma$ is a hyper-parameter balancing energy efficiency and classification performance. As depicted, classification with power control yields desired accuracy levels  with order of magnitudes lower TX power during inference. \tochange{This behavior exemplifies this MINN application for low and varying SNR levels.}}
    \label{fig:power-control}
\end{figure}

To promote energy efficient MINN implementations, their data-driven design objective may be augmented with a penalty term including the TX power at the {\em Encoder} module. This practice is similar to DNN regularization and can also be interpreted as MINN optimization under soft constraints. More specifically, a scalar parameter may be used to control the magnitude of the penalty term as the MINN progressively reduces the transmission power.
As depicted in Fig.~\ref{fig:power-control}, MINNs with integrated power control achieve comparable classification performance with orders of magnitude lower final power budget, as compared with MINNs devoiding this control.

\begin{figure}[t]
    \centering
    \includegraphics[width=\linewidth]{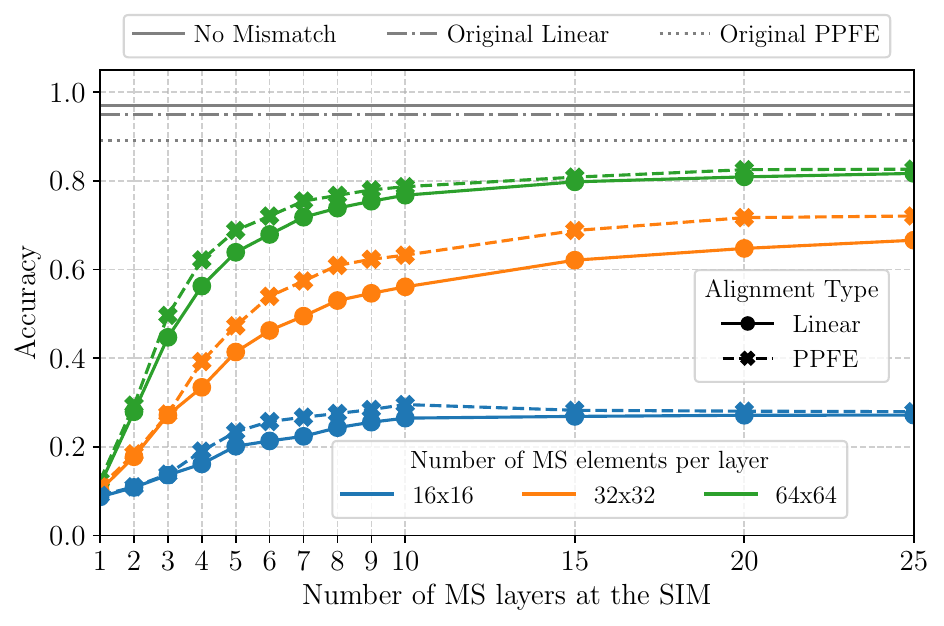}
    \caption{\tochange{Mean accuracy of different MINN versions for the objective of OTA alignment for MNIST classification. Two pairs of {\em Encoder}/{\em Decoder} modules, corresponding to two distinct TX/RX MIMO links for EI, were first pretrained independently from each other to encode and classify CIFAR-10 images (i.e., learned different encoded spaces that were semantically misaligned). Then, the {\em Channel} module was optimized to align those encoded spaces, enabling efficient OTA classification between the {\em Encoder} module of the TX of the first pair and the {\em Decoder} of the RX of the second pair. This process involved learning a linear transformation that maps the encoding space of the first MIMO pair to the encoding space of the second one. The optimal mapping was first computed digitally, either as a direct linear transformation or by applying Proto Parseval Frame Equalizers (PPFE) onto the encoded data as a pre-processing step. Then, the {\em Channel} module controlling a SIM placed at the TX, with different numbers of layers and metamaterials, was optimized to approximate the ideal transformation matrices. The all-MSs MIMO variation of the MINN architecture in Fig.~\ref{fig:MS-positions}(a) was considered, with the TX's SIM having a first layer of $768$ metamaterials whose responses were designed to correspond to the actual image values, and the RX was equipped with a $384$-element fully absorbing MS comprising the first DNN layer of the {\em Decoder}. As observed, when the number of SIM layers and their elements increase, the proposed OTA semantic alignment's performance approaches that of the simulated fully digital alignment techniques. 
    }}
    \label{fig:sim-alignment}
\end{figure}

\tochange{A recent research direction deals with SIM response optimization to approximate arbitrary matrices~\cite{simDFT}. Inspired by this MS-based operation approximation potential, a fully digital DNN may first be trained to perform EI and, subsequently, replace one of their hidden layers with an appropriately designed MS or multi-layer MS structure~\cite{Gunduz_Layer_Approximation}. Under the MINN framework, this approach indicates that the original digital weights may be approximated by the programmable component of the {\em Channel} module, possibly in conjunction with the TX beamformer ({\em Encoder} module) and RX combiner ({\em Decoder} module). An application of this idea for the objective of OTA semantic (i.e., encoding) alignment~\cite{paolo_semantic_alignment_SIM} is presented in Fig.~\ref{fig:sim-alignment}. As shown, MINNs employing deeper and larger SIM within the programmable {\em Channel} module achieve performance comparable to fully digital alignment techniques.}

\section{Open Challenges and Future Directions}\label{sec:challenges}
Despite the recent realizations of the MINN framework, capitalizing on state-of-the-art D\textsuperscript{2}NN prototypes and SIM architectures, there exist several open challenges ranging from hardware design, algorithmic development, and prototyping. 

\tochange{\textbf{Hardware Components and Orchestration:}}
\tochange{Practical MS designs implement quantized responses, necessitating non-negligible power to maintain and change their configurations as well as to operate their controllers, which, for the case of training for dynamic responses, need to also support a DNN. For this case, MSs with integrated sensing and computing capabilities~\cite{AlexandropoulosRIS} may provide attractive solutions for locally acquiring channel knowledge. However, the impact of quantization on training has not been yet explored, and efficient orchestration protocols are needed for the distribution of the trained parameters at all three MINN modules. It is crucial for the proposed {\em Encoder} and {\em Decoder} modules to operate under reasonable power consumption supporting both ML computations and conventional communication functions. For the former objective, multi-hybrid XL MIMO solutions with few and low power active components are needed (small numbers of RF chains with low resolution signal converters). Finally, the placement of the MS-based DNN layers within the signal propagation environment and/or near the end devices is of paramount importance to ensure proper illumination of the MSs during both the forward and backward network passes.}

\tochange{\textbf{Nonlinear Analog-/Wave-Domain Layers:}}
\tochange{D\textsuperscript{2}NNs and recent hybrid digital-/wave-domain DNN designs~\cite{Stylianopoulos_GO,paolo_semantic_alignment_SIM} rely on linear response MSs, implying that MINN's {\em Channel} module behaves overall as a single linear layer, irrespective of the number of SIM layers incorporated. This feature, however,} provides only limited approximation capabilities. To enable universal OTA approximation, nonlinear activation functions must be integrated in each layer-to-layer OTA/analog connection, which necessitates novel designs of metamaterials and/or passive/near-passive RF circuits. Note that nonlinear responses need not be controllable, since DNN activation functions are typically fixed. \tochange{For this goal, RF components operating in their saturation region~\cite{Stylianopoulos_MIMO_ELM} and diode-based RF circuits approximating the Rectified Linear Unit (ReLU) activation~\cite{NL_ELM} constitute promising research directions.}

\textbf{Advanced DNN Architectures:}
State-of-the-art ML models employ more advanced layer architectures, going beyond what is currently offered by the fully-connected feedforward propagation of D\textsuperscript{2}NNs.
Encouraging attempts have been recently made to implement convolutional layers by exploiting wideband characteristics~\cite{AirNN}, however, implementing deep convolutional neural networks, or more advanced recurrent and attention-based architectures poses a formidable challenge.

\textbf{Theoretical Inference Guarantees:}
Ensuring the universal approximation properties of MINN's {\em Channel} module is a tedious task. In fact, further theoretical advancements are required to extend the guarantees of MINN-ELMs~\cite{Stylianopoulos_MIMO_ELM} to deeper structures and dynamic fading scenarios. \tochange{To this end, accurate MSs-parametrized channel modeling is crucial, impacting also simulations-based offline training.} Complementary, analytical insights may guide 
regularization, initialization, and hyper-parameter selection, offering more stable training behavior.

\tochange{\textbf{MINNs with Wideband Signaling:}
D\textsuperscript{2}NNs mainly rely on the spatial degrees of freedom offered by multiple antenna/MS elements, with the temporal and frequency dimensions having been less explored~\cite{LML23_D2NN,Luo2019_Wideband_D2QN}. Wideband MS designs~\cite{AlexandropoulosRIS} and signaling can enable extensive parallel data transmissions in MINNs, facilitating feature extraction with minimal digital processing at the endpoints~\cite{LAY26_wideband_SIM}. Besides, empowering MINNs with temporal memory, going beyond linear time invariant systems, may serve as a means to implement recurrent layers.}


\textbf{OTA MINN Training:}
The training of all MINN modules 
can be performed with synthetic data on realistic simulators. This implies that gradient updates can be computed digitally and, then, the optimized responses applied to the respective MS layers of the E2E, possibly heterogeneous, DNN. It is, however, desirable to design multi-layer MS structures and D\textsuperscript{2}NN enabling OTA calculation of objective functions and their corresponding gradients. In this way, reconfiguring MS responses based on impinging error signals will allow for wave-domain-based backward passes.

\textbf{Distributed MINNs:}
The MINN framework may be realized across multiple devices with heterogeneous characteristics, shaping scalable PHY-layer-based DNNs on demand. This distributed OTA computing necessitates the development of new orchestration schemes and protocols for training and inference. Especially, when extending the MINN framework to multi-D\textsuperscript{2}NN/-user and federated learning scenarios, low overhead synchronization schemes are of particular importance.

\section{Conclusion}\label{sec:conclusion}
\tochange{This article presented the concept of MINNs, a PHY-layer-enabled heterogeneous DNN framework encompassing layers realized in the digital, analog, and wave-propagation domains, offering OTA ML applications in a computations placement flexible manner for future edge networking with lightweight devices. The operation of the three constituent modules of the proposed E2E MIMO system, all primarily comprising multi-layer MS structures and/or SIM, was discussed, followed by variations of the overall architecture and applications for the example of image classification. The framework's training mechanism for both static and dynamic MS responses was also elaborated. Finally, a list of MINN open challenges and respective research directions was presented.}

\balance

\bibliographystyle{IEEEtran}
\bibliography{references}

@article{XYN18_Diffractive_DNN,
author = {Xing Lin and Yair Rivenson and Nezih T. Yardimci and Muhammed Veli and Yi Luo and Mona Jarrahi and Aydogan Ozcan},
title = {All-optical machine learning using diffractive deep neural networks},
journal = {Science},
volume = {361},
number = {6406},
pages = {1004-1008},
year = {2018},
}

@Article{LML23_D2NN,
author={Liu, Che and Ma, Qian and Luo, Zhang Jie and Hong, Qiao Ru and Xiao, Qiang and Zhang, Hao Chi and Miao, Long and Yu, Wen Ming and Cheng, Qiang and Li, Lianlin and Cui, Tie Jun},
title={A programmable diffractive deep neural network based on a digital-coding metasurface array},
journal={Nat. Electron.},
year={2022},
day={01},
volume={5},
number={2},
pages={113-122}
}

@article{Stylianopoulos_GO,
    author = {Stylianopoulos, Kyriakos and {Di Lorenzo}, Paolo and Alexandropoulos, George C.},
    title = {Over-the-Air Edge Inference via Metasurfaces-Integrated Artificial Neural Networks},
    journal = {arXiv preprint arXiv:2504.00233},
    year = {2025}
}

@article{Stylianopoulos_MIMO_ELM,
    author = {Stylianopoulos, K. and Alexandropoulos, G. C.},
    title = {Universal Approximation with {XL} {MIMO} Systems: {OTA} Classification via Trainable Analog Combining},
    journal = {arXiv preprint arXiv:2504.12758},
    year = {2025}
}

@ARTICLE{AXN23_SIM,
  author={An, Jiancheng and Xu, Chao and Ng, Derrick Wing Kwan and Alexandropoulos, George C. and Huang, Chongwen and Yuen, Chau and Hanzo, Lajos},
  journal={IEEE J. Sel. Areas Commun.}, 
  title={Stacked Intelligent Metasurfaces for Efficient Holographic {MIMO} Communications in {6G}}, 
  year={2023},
  volume={41},
  number={8},
  pages={2380-2396},
}

@article{GJZ24_SIM_TOC,
  author={Huang, Guojun and An, Jiancheng and Yang, Zhaohui and Gan, Lu and Bennis, Mehdi and Debbah, Mérouane},
  journal={IEEE Wireless Commun. Lett.}, 
  title={Stacked Intelligent Metasurfaces for Task-Oriented Semantic Communications}, 
  year={2025},
  volume={14},
  number={2},
  pages={310-314}
}

@article{paolo_semantic_alignment_SIM,
  author={M. E. Pandolfo and K. Stylianopoulos and G. C. Alexandropoulos and Di Lorenzo, P.},
  journal={arXiv preprint arXiv:2512.05657},
  title={Over-the-air semantic alignment with stacked intelligent metasurfaces},
  year={2026},
}

@book{AlexandropoulosRIS,
  author    = {Alexandropoulos, G. C. and Zappone, A. and Shlezinger, N. and Di Renzo, M. and Eldar, Y. C.},
  title     = {Reconfigurable Intelligent Surfaces for Wireless Communications: Modeling, Architectures, and Applications},
  publisher = {Springer Nature},
  address   = {Singapore},
  year      = {2026}
}

@article{Gunduz_Layer_Approximation,
      title={Implementing Neural Networks Over-the-Air via Reconfigurable Intelligent Surfaces}, 
      author={Meng Hua and Chenghong Bian and Haotian Wu and Deniz Gündüz},
      year={2025},
    journal={arXiv preprint arXiv:2508.01840},
}

@ARTICLE{simDFT,
  author={An, Jiancheng and Yuen, Chau and Guan, Yong Liang and Di Renzo, Marco and Debbah, Mérouane and Poor, H. Vincent and Hanzo, Lajos},
  journal={IEEE J. Sel. Areas Commun.}, 
  title={Two-Dimensional Direction-of-Arrival Estimation Using Stacked Intelligent Metasurfaces}, 
  year={2024},
  volume={42},
  number={10},
  pages={2786-2802}
}

@Article{Luo2019_Wideband_D2QN,
author={Luo, Yi and Mengu, Deniz and Yardimci, Nezih T. and Rivenson, Yair and Veli, Muhammed and Jarrahi, Mona and Ozcan, Aydogan},
title={Design of task-specific optical systems using broadband diffractive neural networks},
journal={Light: Sci. Appl.},
year={2019},
day={02},
volume={8},
number={1},
pages={112},
}

@ARTICLE{AirComp_Review,
  author={Wang, Zhibin and Zhao, Yapeng and Zhou, Yong and Shi, Yuanming and Jiang, Chunxiao and Letaief, Khaled B.},
  journal={IEEE Internet Things J.}, 
  title={Over-the-Air Computation for {6G}: {F}oundations, Technologies, and Applications}, 
  year={2024},
  volume={11},
  number={14},
  pages={24634-24658},
}

@ARTICLE{AirNN,
  author={Garcia Sanchez, Sara and Reus-Muns, Guillem and Bocanegra, Carlos and Li, Yanyu and Muncuk, Ufuk and Naderi, Yousof and Wang, Yanzhi and Ioannidis, Stratis and Chowdhury, Kaushik Roy},
  journal={IEEE/ACM Trans. Netw.}, 
  title={{AirNN}: {O}ver-the-Air Computation for Neural Networks via Reconfigurable Intelligent Surfaces}, 
  year={2023},
  volume={31},
  number={6},
  pages={2470-2482},
  }

@inproceedings{NL_ELM,
    author = {Kyriakos Stylianopoulos and Mattia Fabiani and Giulia Torcolacci and Davide Dardari and George C. Alexandropoulos},
    title = {Over-The-Air Extreme Learning Machines with {XL} Reception via Nonlinear Cascaded Metasurfaces},
    booktitle={Proc. Int. Zurich Seminar Inf. Commun.},
    year={2026 (arXiv preprint arXiv:2601.17749)},
    address={Zurich, Switzerland},
}

@ARTICLE{LAY26_wideband_SIM,
  author={Li, Zheao and An, Jiancheng and Yuen, Chau},
  journal={IEEE Trans. Wireless Commun.}, 
  title={Stacked Intelligent Metasurface-Enhanced {MIMO} {OFDM} Wideband Communication Systems}, 
  year={2026},
  volume={25},
  number={},
  pages={9608-9622},
 }

\end{document}